\documentclass[preprint,preprintnumbers,nofootinbib]{revtex4}

\usepackage[dvips]{color}
\usepackage{graphicx}
\usepackage{amsmath,amssymb}

\newcommand{\hif}{\mathchar`-}

\newcommand{\ev}{{\rm eV}}

\newcommand{\bmx}{\left(\begin{array}}
\newcommand{\emx}{\end{array}\right)}
\newcommand{\eqn}[1]{&\hspace{-0.2em}#1\hspace{-0.2em}&}

\begin{document}
\vspace{2cm}
\preprint{MISC-2012-22}

\title{Partial mass degeneracy and spontaneous CP violation in the lepton sector}
\author{
Takeshi Araki\footnote{araki@cc.kyoto-su.ac.jp} and Hiroyuki Ishida\footnote{h\_ishida@cc.kyoto-su.ac.jp} } \affiliation{ Maskawa Institute, Kyoto Sangyo University,\\ Motoyama, Kamigamo, Kita-Ku, Kyoto 603-8555, Japan}

\begin{abstract}
Inspired by the small mass-squared difference measured in the solar neutrino oscillation experiments and by the testability, we suggest that a limit of the partial mass degeneracy, in which masses of the first two generation fermions are degenerate, may be a good starting point for understanding the observed fermion mass spectra and mixing patterns.
The limit indicates the existence of a two-dimensional rotation symmetry, such as $O(2)$, $D_N^{}$ and so on, in flavor space of the first two generations.
We propose simple models for the lepton sector based on $D_N^{}$ and show that the models can successfully reproduce the experimental data without imposing unnatural hierarchies among dimensionless couplings, although at least $10\%$ tuning is necessary in order to explain a large atmospheric mixing.
It is especially found that the $Z_2^{}$ subgroup of the $D_N^{}$ symmetry plays an important role in understanding the smallness of the electron mass and $\theta_{13}^{\rm PMNS}$.
We also discuss testability of the models by the future neutrinoless-double-beta-decay experiments and cosmological observations.
\end{abstract}

\maketitle

\section{Introduction and Motivation}
Although the Standard Model (SM) of particle physics agrees very well with various experimental results, some problems and unsatisfactory points have also been pointed out.
One of them is the lack of a guiding principle ruling the flavor structure of fermions.
In this regard, the introduction of flavor symmetries is one of the most conceivable extensions of the SM 
and also well motivated from a viewpoint of string theories \cite{stringy}.

Since, as concluded in Ref.~\cite{no-go}, most of flavor symmetries need to be broken either spontaneously or explicitly at energy scales much above the electroweak scale, it is usually not easy to decide which symmetries are suitable and how they should be imposed.
To this end, one needs to find out remnants of such broken symmetries in a low energy Lagrangian, and small parameters could be important for this purpose.
If symmetry breaking was slight, its effects should correspondingly be small, yielding small breaking terms in a low energy Lagrangian.
Moreover, even if they were grossly broken, the breaking effects might be realized as effective non-renormalizable operators suppressed by their breaking scales.
For instance, it is well known that a Majorana neutrino mass term can be constructed with only the SM particles at mass-dimension five \cite{dim5}:
\begin{eqnarray}
{\cal L}_{\rm eff}=\frac{f_{ij}}{\Lambda_\nu}L_iL_jHH,
\label{eq:wein}
\end{eqnarray}
where $L$ and $H$ represent the left-handed lepton and the SM Higgs doublets, respectively, and that the lepton number conservation is
violated with this term\footnote{In Ref. \cite{chen}, authors relate the tiny neutrino masses with the Peccei-Quinn symmetry.}.
In this case, this term can be regarded as a breaking term of the lepton number symmetry broken at the high energy scale $\Lambda_\nu$.
Hence, it might be said that small parameters in a low energy theory are the manifestation of new symmetries in high energies; zero limits of the small parameters may correspond to the unbroken limits of the associated symmetries.

In the case of flavor symmetries, their remnants should appear in the fermion mass terms and/or the flavor mixing sectors. 
For example, one can line up the following candidates of small parameters:
\begin{enumerate}
\item $\theta_{13}^{\rm PMNS}\ll\theta_{12}^{\rm PMNS}, ~\theta_{23}^{\rm PMNS}$ ~~~({\rm or}~~$|V^{\rm PMNS}_{e3}|\ll {\rm the~others}$),
\item $|\theta_{23}^{\rm PMNS}-45^\circ|\ll \theta_{23}^{\rm PMNS}$~~~({\rm or}~~$||V^{\rm PMNS}_{\mu 3}|-1/\sqrt{2}|\ll |V^{\rm PMNS}_{\mu 3}|$),
\item $\Delta m^2_{12}=(m_2^\nu)^2 - (m_1^\nu)^2 \ll \Delta m^2_{23} =|(m_3^\nu)^2 - (m_2^\nu)^2|$,
\item $m_1^{u,d,\ell},~m_2^{u,d,\ell} \ll m_3^{u,d,\ell}$,
\item $\theta_{ij}^{\rm CKM} \ll \theta_{ij}^{\rm PMNS}$~~~({\rm or}~~$|V^{\rm CKM}_{ij}| \ll |V^{\rm PMNS}_{ij}|$),
\end{enumerate}
where $\theta_{ij}^{\rm PMNS}$ and $\theta_{ij}^{\rm CKM}$ stand for the mixing angles of the Pontecorvo-Maki-Nakagawa-Sakata (PMNS), $V^{\rm PMNS}$, and the Cabibbo-Kobayashi-Maskawa (CKM), $V^{\rm CKM}$, mixing matrices\footnote{In this paper, we adopt the standard parametrization \cite{PDG}.}, respectively;
$m_{i}^{f}$ with $f=u,d,\ell,\nu$ and $i=1,2,3$ denotes masses of the up-type quarks, down-type quarks, charged leptons and neutrinos.
The first two items have attracted a lot of attention over the years as they predict the $\mu\hif\tau$ permutation symmetry \cite{mutau} or non-Abelian discrete flavor symmetries \cite{nADFS} in the limit of $\theta_{13}^{\rm PMNS}=0^\circ$ with $\theta_{23}^{\rm PMNS}=45^\circ$.
In fact a number of models based on these symmetries have been proposed \cite{models}.
The recent reactor \cite{rct} and long-baseline \cite{acl} neutrino oscillation experiments, however, disfavor the vanishing $\theta_{13}^{\rm PMNS}$: the DAYA-BAY experiment \cite{8sigma} give us $7.7\sigma$ deviations from $\theta_{13}^{\rm PMNS}=0^\circ$ with a rather large central value of $\theta_{13}^{\rm PMNS}\simeq 8.7^\circ$.
Furthermore, considerable deviations of $\theta_{23}^{\rm PMNS}$ from $45^\circ$ have also been found in the global analysis of the neutrino oscillation experiments \cite{valle, fogli, garc}, yet its confidence level is not high enough to conclude.
Having these facts in mind, we would turn our attention to the remaining three items and discuss their consequences in the present study.

First, the third item means quasi mass degeneracy between the first and second generation neutrinos unless $m_1^\nu$ is much smaller than $m_2^\nu$ in the case of normal mass ordering.
Let us focus on this partially quasi-degenerated region (roughly $m_1^\nu = 0.05 \sim 0.1~\ev$ and $m_3^\nu=0 \sim 0.1~\ev$ for the normal and inverted ordering cases, respectively) and consider the effective Majorana neutrino mass operator given in Eq. (\ref{eq:wein}).
In the limit of $\Delta m_{12}^2=0$, the Majorana neutrino mass matrix comes to respect a two-dimensional rotation symmetry \cite{o2,general}\footnote{
The mass degeneracy is entangled with the mixing in Ref. \cite{vonDyck}.}, such as $O(2)$, $SO(2)$ and $D_N^{}$:
\begin{eqnarray}
&&
R^T
\bmx{ccc}
 m_1^\nu & 0 & 0 \\
 0 & m_1^\nu & 0 \\
 0 & 0 & m_3^\nu
\emx
R
=
\bmx{ccc}
 m_1^\nu & 0 & 0 \\
 0 & m_1^\nu & 0 \\
 0 & 0 & m_3^\nu
\emx
~~~{\rm with}~~~
R=
\bmx{ccc}
 \cos\theta & \sin\theta & 0 \\
-\sin\theta & \cos\theta & 0 \\
 0 & 0 & 1
\emx.
\label{eq:Mnu0}
\end{eqnarray}
In other words, $L_1$ and $L_2$ belong to a doublet representation, e.g. ${\bf 2}_n$, of the symmetry, while $L_3$ behaves as a singlet representation, e.g. ${\bf 1}$. 
(Notations of $O(2)$ are given in Appendix.)
Then, the observed slight mass splitting between $m_1^\nu$ and $m_2^\nu$ could be interpreted as slight breaking of the rotation symmetry.
Like this, the observed neutrino mass spectrum could be explained by starting from a limit of $m_1^\nu = m_2^\nu$.

Secondly, the idea of the partial mass degeneracy ($m_1^f=m_2^f$), however, seems to conflict with the fourth item because the charged lepton masses are strongly hierarchical.
Nevertheless, it is possible to realize a hierarchical mass spectrum in the case of Dirac fermions by assigning a different doublet representation, ${\bf 2}_{m\neq n}$, or a singlet representation ${\bf 1}$ to the right-handed charged leptons, resulting in
\begin{eqnarray}
M^{\ell}=
\bmx{ccc}
0 & 0 & 0 \\
0 & 0 & 0 \\
0 & 0 & m_{33}^\ell
\emx
~~~~{\rm or}~~~~
M^\ell =
\bmx{ccc}
0 & 0 & 0 \\
0 & 0 & 0 \\
m_{31}^{\ell} & m_{32}^{\ell} & m_{33}^{\ell}
\emx ,\label{eq:Me0}
\end{eqnarray}
respectively.
It can be readily observed that the electron and muon masses are vanishing and thus degenerate in both cases.
In this sense, the idea of the partial mass degeneracy may be applicable to not only the charged lepton sector but also the quark sectors.
Rather, that appears reasonable since one can relate the smallness of the light charged fermion masses with symmetry breaking.
Also, it may enable us to naturally understand some phenomenological relations among elements of the CKM matrix \cite{xing}.

Lastly, given the above conjectures, the fifth item may naturally be explained at the same time.
Breaking of the rotation symmetry triggers flavor mixing as well. 
On one hand, in the quark and charged lepton sectors, small flavor mixings are expected because breaking terms are supposed to be small; thus the observed small CKM mixing would be derived\footnote{Note that mixing between the first and second generations is not necessarily small because of the mass degeneracy. This might enable us to understand why $\theta_{12}^{\rm CKM}$ is a little larger than the others.}.
On the other hand, in the neutrino sector, its flavor mixing can be large since the leading-order neutrino mass matrix in Eq. (\ref{eq:Mnu0}) is almost proportional to the unit matrix in the neutrino mass regions under consideration.
We here stress that the mysterious differences between the CKM and PMNS matrices stem from the nature of fermions, that is to say Dirac or Majorana, in this scenario.

To summarize, the limit of the partial mass degeneracy seems to fit the observed fermion mass spectra and mixing patterns and suggests the existence of a two-dimensional rotation symmetry.
Yet another motivation to consider the partial mass degeneracy is that the effective mass, $\langle m_\nu^{} \rangle$, of the neutrinoless double beta decay is not vanishing even in the case of normal mass ordering\footnote{We would like to thank E. Takasugi for making us aware of this point.}.
In Fig. \ref{fig:0nbb}, we depict the allowed regions of $\langle m_\nu^{} \rangle$ as a function of the lightest neutrino mass in the standard $3\nu$ framework, where 3$\sigma$ constraints of the neutrino oscillation parameters from \cite{fogli} are imposed while varying CP phases from $0$ to $2\pi$.
For example, if $m_1^\nu > 0.05~\ev$, then $\langle m_\nu^{} \rangle > 0.01~\ev$, which would be accessible by the next generation EXO and KamLAND-Zen experiments \cite{neut12}.
In addition, the sensitivity of cosmic microwave background (CMB) observations on the neutrino mass has just started to enter this mass region \cite{planck}.
Thus, it should be worthwhile to carry out theoretical studies on this mass region.

In Sect. II, we show a simple model for the lepton sector by means of a $D_N^{}$ flavor symmetry and demonstrate that its $Z_2^{}$ subgroup forbids the electron mass and $\theta_{13}^{\rm PMNS}$.
The particle content is enriched with SM-gauge-singlet real scalars.
In order to break the $Z_2^{}$ symmetry, in Sect. III, we promote the singlet scalars to complex ones with complex vacuum expectation values.
After mentioning testability of the models in Sect. IV, we summarize our results and discuss what to do next in Sect. V.
The group theories of $D_N^{}$ and $O(2)$ are briefly summarized in Appendix A and B, respectively.

\section{Model with real scalars}
\begin{table}[ht]
\begin{tabular}{|c|c|c|c|c|c|}\hline
 & $L_I$ & $L_3$ & $\ell_{i}$ & $H$ & $S_I$ \\ \hline
$D_N$ & ${\bf 2}_2$ & ${\bf 1}$ & ${\bf 1}$ & ${\bf 1}$ & ${\bf 2}_1$ \\ \hline
\end{tabular}
\caption{A particle content and charge assignment of the model, where $I=1,2$ and $i=1\hif 3$ denote the indices of generations; 
$L$ and $\ell$ represent the left- and right-handed SM leptons, respectively; $H$ and $S_{1,2}^{}$ are the SM Higgs and gauge-singlet real scalars, respectively.}
\label{tab:1}
\end{table}

Typical examples of a flavor symmetry realizing the partial mass degeneracy are presumably $O(2)$, $SO(2)$ and $D_N^{}$.
As we shall explain later, $SO(2)$ may be excluded from the list since it does not include a $Z_2$ parity, whereas both $O(2)$ and $D_N^{}$ would be useful for our purpose.
We here adopt $D_N^{}$ as our flavor symmetry just to avoid dangerous massless Nambu-Goldstone bosons or gauge anomalies.
We concentrate on the case of $N={\rm odd}$ and postulate that $N$ is sufficiently large.
In this case, one may be able to disregard the tensor product Eq. (\ref{eq:cg3}) in Appendix A unless higher order terms with huge suppression factors are concerned, and the theory would be governed by only Eqs. (\ref{eq:cg1}) and (\ref{eq:cg2}) which are the same as those of $O(2)$.
In this sense, the following models would work for $O(2)$, too.
We will make a further comment on differences between $D_N^{}$ and $O(2)$ models later in this section.

We embed $L_1^{}$ and $L_2^{}$ into the doublet representation ${\bf 2}_2$ of $D_N^{}$ and assign the trivial singlet representation ${\bf 1}$ to the other leptons and the SM Higgs\footnote{Alternatively, one can embed $\ell_1$ and $\ell_2$ into a doublet representation too as mentioned in Section I. In this case, however, masses of the electron and muon tend to be degenerate within the given particle contents. Such charge assignments are adopted in Ref. \cite{old-models} with different particle contents.}, resulting in $m_e^{}=m_\mu^{}=0$ and $m_1^\nu = m_2^\nu$.
In order to lift the mass degeneracies, the $D_N$ flavor symmetry must be broken by a doublet representation, so that we introduce a set of SM-gauge-singlet real scalars $S^{}_{1,2}$ belonging to ${\bf 2}_1$.
A particle content and charge assignment of the model are summarized in Table \ref{tab:1}.
We consider the effective Majorana neutrino mass operator given in Eq. (\ref{eq:wein}) so as to keep our discussions as general as possible.
Under the $D_N$ flavor symmetry, the charged lepton Yukawa and Majorana neutrino mass terms are written by
\begin{eqnarray}
{\cal L}
&=& 
  y_{i}^{0} ~\overline{L}_3^{} H \ell_i^{}
+ \frac{y^{}_{i}}{\Lambda_F^2} \overline{L}_I^{} H \ell_{i}^{} (S_{}^2)_I^{}
+ \frac{f_\nu}{\Lambda_\nu}L_I L_I H H
+ \frac{f_\nu^\prime}{\Lambda_\nu}L_3 L_3 H H
\nonumber \\
&&
+ \frac{g_\nu}{\Lambda_\nu \Lambda_F^2}L_3 L_I H H (S^2)_I
+ \frac{h_\nu}{\Lambda_\nu \Lambda_F^4}(L_J L_K)_I H H (S^4)_I\,,
\end{eqnarray}
where $y_i^0$, $y_i^{}$, $f_\nu^{}$, $f_\nu^\prime$, $g_\nu^{}$ and $h_\nu^{}$ are dimensionless complex couplings and supposed to be ${\cal O}(1)$, and $\Lambda_F$ describes a breaking scale of the $D_N$ symmetry.
Note that the term proportional to ${\cal O}(1/\Lambda_F^{4})$ is omitted in the charged lepton sector since it is absorbed by $(y_i^{}/\Lambda_F^2) ~\overline{L}_I^{} H \ell_{i}^{} (S_{}^2)_I^{}$.
We denote the vacuum expectation values (VEVs) of the scalars as
\begin{eqnarray}
\langle H \rangle=v,~~
\langle S_I \rangle=(s_1~~s_2)^T, \label{eq:vev_real}
\end{eqnarray}
which yield the following mass matrices for the charged leptons and neutrinos:
\begin{eqnarray}
\frac{1}{v} M^\ell \eqn{=} 
\begin{pmatrix}
0 &0 &0\\
0 &0 &0\\
y_1^0 &y_2^0 &y_3^0
\end{pmatrix}
+ 
\frac{1}{\Lambda_F^2}
\begin{pmatrix}
y_1 \delta_1 &y_2 \delta_1 &y_3 \delta_1 \\
y_1 \delta_2 &y_2 \delta_2 &y_3 \delta_2 \\
0 &0 &0
\end{pmatrix}\,,
\label{eq:Me1}
\end{eqnarray}
\begin{eqnarray}
\frac{\Lambda_\nu}{v^2} M^\nu 
= 
\begin{pmatrix}
f_\nu &0 &0\\
0 &f_\nu &0\\
0 &0 &f_\nu'
\end{pmatrix}
+
\frac{1}{\Lambda_F^2}
\begin{pmatrix}
0 &0 &g_\nu \delta_1 \\
0 &0 &g_\nu \delta_2 \\
g_\nu \delta_1 &g_\nu \delta_2 &0
\end{pmatrix} 
+ 
\frac{1}{\Lambda_F^4}
\begin{pmatrix}
 h_\nu (\delta_1^2 - \delta_2^2) & 
 h_\nu 2\delta_1^{}\delta_2^{} & 0\\
 h_\nu 2\delta_1^{}\delta_2^{} &
-h_\nu (\delta_1^2 - \delta_2^2) & 0\\
0 &0 &0
\end{pmatrix} \,.
\label{eq:Mn1}
\end{eqnarray}
Here and hereafter we use the following abbreviations
\begin{eqnarray}
\delta_1^{} = s_1^2 - s_2^2\,,
\eqn{} \delta_2^{} =2 s_1^{} s_2^{}\,,~
\delta_s^{}=s_1^2+s_2^2\,.
\end{eqnarray}
The charged lepton mass matrix is diagonalized by the unitary transformation
\begin{eqnarray}
V^\ell_{} 
= 
\frac{1}{\delta_s^{}} 
\begin{pmatrix}
-\delta_2 &\delta_1 &0\\
\delta_1 &\delta_2 &0\\
0 &0 & \delta_s^{}
\end{pmatrix}
\begin{pmatrix}
1 &0 &0\\
0 &\cos \theta &-\sin \theta ~e^{i\rho}\\
0 &\sin \theta ~e^{-i\rho} &\cos \theta
\end{pmatrix}\,,
\label{eq:Ve1}
\end{eqnarray}
with
\begin{eqnarray}
\tan 2 \theta = \frac{2 Y' \delta_s^{} \Lambda_F^2}{Y \delta_s^2  - Y^0 \Lambda_F^4}
\sim {\cal O}\left(\frac{1}{\Lambda_F^2}\right)\,,
\end{eqnarray}
where $Y^0 = |y_1^0|^2 + |y_2^0|^2 + |y_3^0|^2$, $Y = |y_1|^2 + |y_2|^2 + |y_3|^2$, $Y' = |y_1^* y_1^0 + y_2^* y_2^0 + y_3^* y_3^0|$ and $\rho={\rm Arg}[y_1^* y_1^0 + y_2^* y_2^0 + y_3^* y_3^0]$.
The eigenvalues are approximately derived as
\begin{eqnarray}
\frac{1}{v^2} (V^\ell_{})^\dag M^\ell (M^{\ell})^\dagger V^\ell \simeq
\begin{pmatrix}
0 &0 &0\\
0 &\frac{Y Y^0 - Y'^2}{Y^0} \frac{\delta_s^2}{\Lambda_F^4} &0\\
0 &0 &Y^0\\
\end{pmatrix}\,.\label{eq:Diag_Ml_R}
\end{eqnarray}
One immediately observes that the electron remains massless and that $s_i^{}/\Lambda_F^{} \sim \sqrt{m_\mu^{}/m_\tau^{}}$.
On the other hand, the unitary transformation affects the neutrino mass matrix in such a way that
\begin{eqnarray}
\frac{\Lambda_\nu}{v^2} 
(V^\ell_{})^T M^\nu V^\ell_{}
\eqn{=} 
\begin{pmatrix}
\bar{M}^\nu_{11} &0 &0\\
0 & \bar{M}^\nu_{22} & \bar{M}^\nu_{23}\\
0 & \bar{M}^\nu_{32} & \bar{M}^\nu_{33} 
\end{pmatrix}\,,
\label{eq:Mnd}
\end{eqnarray}
where
\begin{eqnarray}
\bar{M}^\nu_{11} \eqn{=} 
f_\nu - h_\nu \frac{\delta_s^2}{\Lambda_F^4}\,,\\
\bar{M}^\nu_{22} \eqn{\sim} 
f_\nu + h_\nu \frac{\delta_s^2}{\Lambda_F^4} 
+ \frac{Y' ( Y' f_\nu' - 2 Y^0 g_\nu)}{(Y^0)^2} \frac{\delta_s^2}{\Lambda_F^4}\,,\\
\bar{M}^\nu_{23} \eqn{=}
\bar{M}^\nu_{32} \sim 
\frac{Y^0 g_\nu + (f_\nu - f_\nu') Y'}{(Y^0)^2} \frac{\delta_s^{} }{\Lambda_F^2}\,,\\
\bar{M}^\nu_{33} \eqn{\sim} 
f_\nu' 
+ \frac{Y' ( Y' f_\nu + 2 Y^0 g_\nu)}{(Y^0)^2} \frac{\delta_s^2}{\Lambda_F^4}\,,
\end{eqnarray}
and it can be seen that $\theta_{12}^{\rm PMNS}$ and $\theta_{13}^{\rm PMNS}$ are vanishing as well.

The vanishing electron mass, $\theta_{12}^{\rm PMNS}$ and $\theta_{13}^{\rm PMNS}$ are not accidental.
In order to explain this, let us first consider the case of $O(2)$.
Since $O(2)$ is a continuous symmetry, 
there always exists an $O(2)$ transformation which keeps the VEV configuration Eq. (\ref{eq:vev_real}) invariant, such as                                                                                                                                                                                                                                                                                                                                     
\begin{eqnarray}
\bmx{cc}
 \cos\theta & \sin\theta \\
 \sin\theta &-\cos\theta
\emx
=\frac{1}{s_1^2 + s_2^2}
\bmx{cc}
 s_1^2 - s_2^2 & 2s_1^{}s_2^{} \\
 2s_1^{}s_2^{} & -(s_1^2 - s_2^2) 
\emx\,.
\label{eq:z2}
\end{eqnarray}
After the symmetry breaking, this invariance ends up an unbroken $Z_2^{}$ symmetry under which the left-handed leptons
transform as
\begin{eqnarray}
L_i^{} \rightarrow
\frac{1}{\delta_s^2}
\bmx{ccc}
 \delta_1^2 - \delta_2^2 & 2\delta_1^{}\delta_2^{} & 0 \\
 2\delta_1^{}\delta_2^{} & -(\delta_1^2 - \delta_2^2) & 0 \\
 0 & 0 & \delta_s^2
\emx_{ij}^{} L_j^{} ,
\label{eq:z2f}
\end{eqnarray}
or in the diagonal basis of the charged lepton mass matrix, it becomes
\begin{eqnarray}
L_i^{} \rightarrow
\bmx{ccc}
 -1 & 0 & 0 \\
 0 & 1 & 0 \\
 0 & 0 & 1
\emx_{ij}^{} L_j^{} \,.
\label{eq:z2m}
\end{eqnarray}
From Eq. (\ref{eq:z2m}), it is clear that the electron mass, $\theta_{12}^{\rm PMNS}$ and $\theta_{13}^{\rm PMNS}$ are forbidden by this $Z_2^{}$ symmetry.
Also, looking back at Eqs. (\ref{eq:Me1}) and (\ref{eq:Mn1}), it can be found that they take the most-general $Z_2$-invariant forms.
In the case of $D_N$, the $Z_2$ symmetry may not be an exact one because $D_N$ is a discrete group.
Corrections stemming from this violation, however, are negligibly small as long as the order of $D_N$ is sufficiently large.
In the case of $N=9$, for instance, effects of the $Z_2$ breaking appear in $(x/\Lambda_F^7) ~\overline{L}_I^{} H \ell_{i}^{} (S_{}^7)_I^{}$ for the first time.
A $D_9^{}$ invariant term is obtained by constructing ${\bf 2}_2$ from $(S^7)_I$ via ${\bf 2}_3 \otimes {\bf 2}_4$ with the tensor product Eq. (\ref{eq:cg3}).
After $S_{1,2}^{}$ develop the VEVs, the resultant term violates the $Z_2^{}$ symmetry and induces a non-zero but negligibly small electron mass.
Similarly, the breaking effects in the neutrino sector are ignorable.
If $N \le 7$, it may be possible to reproduce a realistic electron mass.
Although this may be an interesting idea, we will discuss a different breaking mechanism of $Z_2^{}$ in the next section while assuming $N \ge 9$.
Note that the $Z_2^{}$ symmetry originates in the parity included in $D_N^{}$ and $O(2)$, whereas $SO(2)$ does not include it.

We stress that the $Z_2^{}$ symmetry could provide us with a natural explanation for the smallness of the electron mass and $\theta_{13}^{\rm PMNS}$: their smallness could be explained by approximate conservation of the $Z_2^{}$ symmetry.
Moreover, the electron mass and $\theta_{13}^{\rm PMNS}$ are possibly correlated with each other through a mechanism of the $Z_2^{}$ symmetry breaking.
In contrast, $\theta_{12}^{\rm PMNS}$ can easily be large due to the mass degeneracy.
In the next section, we will show it is indeed the case.

\section{Model with complex scalars}
\subsection{Model}
One way to break the $Z_2^{}$ symmetry is to promote $S_{I}^{}$ to complex scalars with complex VEVs. 
In the current and next subsections, just for simplicity, we assume that all the dimensionless couplings are real and invoke spontaneous CP violation (SCPV).
The availability of SCPV will be discussed in the next subsection, and here we simply rewrite the VEVs of the scalars as
\begin{eqnarray}
\langle H \rangle=v,~~
\langle S_I \rangle=(s_1 e^{i\phi_1}~~s_2 e^{i\phi_2})^T\,.\label{eq:vev}
\end{eqnarray}
As can be seen from Eq. (\ref{eq:z2}), an unbroken limit of the $Z_2^{}$ symmetry corresponds to $\phi_1^{} = \phi_2^{}$, which keeps the right-hand side real, and thus a slight splitting between them is expected to trigger a non-zero electron mass, $\theta_{12}^{\rm PMNS}$ and $\theta_{13}^{\rm PMNS}$.
The Lagrangian is augmented by the following new terms:
\begin{eqnarray}
{\cal L}_{\rm new}^{}
&=&\frac{y^{'}_{i}}{\Lambda_F^2} \overline{L}_I^{} H \ell_{i}^{} (S_{}^{*2})_I
+ \frac{y''_{i}}{\Lambda_F^2} \overline{L}_I^{} H \ell_{i}^{} (|S_{}|^2)_I 
\nonumber \\
&&
+ \frac{g_\nu^\prime}{\Lambda_\nu \Lambda_F^2}L_3 L_I H H (S^{*2})_I
+ \frac{g_\nu''}{\Lambda_\nu \Lambda_F^2}L_3 L_I H H (|S|^2)_I 
\nonumber \\
&&
+ \frac{h_\nu^\prime}{\Lambda_\nu \Lambda_F^4}(L_J L_K)_I H H (S^{*4})_I
+ \frac{h_\nu''}{\Lambda_\nu \Lambda_F^4}(L_J L_K)_I H H (|S|^4)_I 
\nonumber \\
&&
+ \frac{h_\nu'''}{\Lambda_\nu \Lambda_F^4}(L_J L_K)_I H H (S^2|S|^2)_I
+ \frac{h_\nu''''}{\Lambda_\nu \Lambda_F^4}(L_J L_K)_I H H (S^{*2}|S|^2)_I\,.
\end{eqnarray}
Suppose $\phi_2 = \phi_1 + \delta \phi$ and $\delta\phi \ll 1$, the charged lepton mass matrix is approximated as
\begin{eqnarray}
\frac{1}{v}M^\ell 
\eqn{\simeq} 
\bmx{ccc}
0 & 0 & 0 \\
0 & 0 & 0 \\
y_1^0 & y_2^0 & y_3^0
\emx 
+
\frac{1}{\Lambda_F^2}
\bmx{ccc}
\tilde{Y}_1^{} \delta_1^{}  & \tilde{Y}_2^{} \delta_1^{} & \tilde{Y}_3^{} \delta_1^{} \\
\tilde{Y}_1^{} \delta_2^{} & \tilde{Y}_2^{} \delta_2^{} & \tilde{Y}_3^{} \delta_2^{}\\
0 & 0 & 0
\emx
-
\frac{i\delta\phi}{\Lambda_F^2}
\bmx{ccc}
 \tilde{Y}_1^\prime \delta_s^{} & \tilde{Y}_2^\prime \delta_s^{} & \tilde{Y}_3^\prime \delta_s^{} \\
 0 & 0 & 0 \\
0 & 0 & 0
\emx\, 
\label{eq:Me2}
\end{eqnarray}
upto the first order of $\delta\phi$, where
\begin{equation}
\begin{split}
\tilde{Y}_i 
&= y_i e^{2i\phi_1}(1+i\delta \phi) + y_i^\prime e^{-2i\phi_1}(1-i\delta \phi) + y_i''\,, \\
\tilde{Y}_i^\prime 
&= y_i e^{2i\phi_1} - y_i^\prime e^{-2i\phi_1}\,.
\end{split}
\end{equation}
Similarly, the neutrino mass matrix is
\begin{eqnarray}
\frac{\Lambda_\nu}{v^2_{}} M^\nu 
&\simeq &
\begin{pmatrix}
f_\nu &0 &0\\
0 &f_\nu &0\\
0 &0 &f_\nu'
\end{pmatrix}
+ 
\frac{1}{\Lambda_F^2}
\begin{pmatrix}
0 &0 &\tilde{G}_\nu^{} \delta_1^{} \\
0 &0 &\tilde{G}_\nu^{} \delta_2^{} \\
\tilde{G}_\nu^{} \delta_1^{} & \tilde{G}_\nu^{} \delta_2^{} &0
\end{pmatrix}
-
\frac{i\delta\phi}{\Lambda_F^2}
\begin{pmatrix}
0 &0 & \tilde{G}_\nu^\prime \delta_s^{}\\
0 &0 & 0 \\
\tilde{G}_\nu^\prime \delta_s^{} & 0 &0
\end{pmatrix} 
\nonumber \\
&&+ 
\frac{1}{\Lambda_F^4}
\begin{pmatrix}
 \tilde{H}_\nu^{} (\delta_1^2 - \delta_2^2) & 
 \tilde{H}_\nu^{} 2\delta_1^{}\delta_2^{} & 0\\
 \tilde{H}_\nu^{} 2\delta_1^{}\delta_2^{} &
-\tilde{H}_\nu^{} (\delta_1^2 - \delta_2^2) & 0\\
0 &0 &0
\end{pmatrix}
-
\frac{i\delta\phi}{\Lambda_F^4}
\begin{pmatrix}
 \tilde{H}_\nu^\prime \delta_1^{}\delta_s^{} & 
 \tilde{H}_\nu^\prime \delta_2^{}\delta_s^{} & 0\\
 \tilde{H}_\nu^\prime \delta_2^{}\delta_s^{} &
-\tilde{H}_\nu^\prime \delta_1^{}\delta_s^{} & 0\\
0 &0 &0
\end{pmatrix} ,
\label{eq:Mn2}
\end{eqnarray}
with
\begin{equation}
\begin{split}
\tilde{G}_\nu^{} 
&= g_\nu^{} e^{2i\phi_1}(1+i\delta \phi) + g_\nu^\prime  e^{-2i\phi_1}(1-i\delta \phi) 
+ g_\nu'' \,,\\
\tilde{G}_\nu^\prime 
&= g_\nu^{} e^{2i\phi_1} - g_\nu^\prime e^{-2i\phi_1}\,,\\
\tilde{H}_\nu^{}
&= [h_\nu^{}e^{4i\phi_1^{}}(1+2i\delta\phi) 
+ h_\nu^\prime e^{-4i\phi_1^{}}(1-2i\delta\phi) + h_\nu''\\
&\hspace{4cm}
+ h_\nu'''e^{2i\phi_1^{}}(1+i\delta\phi) + h_\nu'''' e^{-2i\phi_1^{}}(1-i\delta\phi)] \,,\\
\tilde{H}_\nu^\prime 
&= [2h_\nu^{}e^{4i\phi_1^{}} - 2h_\nu' e^{-4i\phi_1^{}} 
+ h_\nu''' e^{2i\phi_1^{}} - h_\nu'''' e^{-2i\phi_1^{}}]\,.
\end{split}
\end{equation}
The terms proportional to $\delta\phi$ violate the $Z_2^{}$ symmetry, which indicates that the electron mass and $\theta_{13}^{\rm PMNS}$ are proportional to $\delta\phi$.
In order to demonstrate this, we here simplify the diagonalization of the charged lepton mass matrix by requiring $\sum_i y_i^0 \tilde{Y}_i^* = \sum_i y_i^0 (\tilde{Y}_i^\prime)^* = \sum_i \tilde{Y}_i^{}(\tilde{Y}_i^\prime)^* = 0$.
With this simplification, we regard the third term of Eq. (\ref{eq:Me2}) as small perturbations.
Then, the electron mass is approximately obtained as
\begin{eqnarray}
m_e^2 \simeq (\delta\phi)^2 \sum_i |Y_i^\prime|^2 \frac{\delta_2^2}{\Lambda_F^4} v_{}^2\,,
\end{eqnarray}
while the $12$ and the $13$ element of the neutrino mass matrix gain
\begin{equation}
\begin{split}
\frac{\Lambda_\nu}{v^2_{}} \bar{M}^\nu_{12} 
&\simeq i\delta\phi\frac{\delta_2^{}\delta_s^{}}{\Lambda_F^4} \tilde{H}_\nu^\prime 
+ {\cal O}\left( \frac{(\delta\phi)^2}{\Lambda_F^4} \right)\,,\\
\frac{\Lambda_\nu}{v^2_{}} \bar{M}^\nu_{13} 
&\simeq i\delta\phi\frac{\delta_2^{}}{\Lambda_F^2} \tilde{G}_\nu^\prime 
+ {\cal O}\left( \frac{(\delta\phi)^2}{\Lambda_F^2} \right)\,.
\end{split}
\end{equation}
in the diagonal basis of the charged lepton mass matrix.
The other elements are almost the same as Eq. (\ref{eq:Mnd}).
Now, it is obvious that $\theta_{13}^{\rm PMNS}$ as well as the electron mass are proportional to and suppressed by $\delta\phi$.
The expression of $\theta_{13}^{\rm PMNS}$ can be derived from them, but it is rather complicated.
Hence, we refrain from showing it.

The model contains a sufficient number of parameters to fit the experimental data.
Nevertheless, we would like to emphasize that the model can reproduce experimental data without manipulating the dimensionless couplings hierarchical by hand.
For instance, we find the following parameter spaces ($\sum_i y_i^0 \tilde{Y}_i^* = \sum_i y_i^0 (\tilde{Y}_i^\prime)^* = \sum_i \tilde{Y}_i^{}(\tilde{Y}_i^\prime)^* = 0$ are not placed):
\begin{equation}
\begin{split}
&y^0_1 = y^0_2 = 1.2,~y^0_3 = 1.0,~
y_1 = -y_2 = y_3  = y_1^\prime = y_2^\prime = -y_3^\prime = 0.8,\\
&y_1'' = -y_2'' = -y_3'' = 0.85\sim 0.90,\\
&f_\nu^\prime=1.0,~f_\nu=0.93\sim 0.95,~
g_\nu = g_\nu'' =0.9,~g_\nu^\prime = 0.9\sim 1.3,\\
&h_\nu = h_\nu^\prime = h_\nu'' = -h_\nu''' = -h_\nu'''' = 0.8 \sim 1.3,\\
&\frac{s_{1,2}}{\Lambda_F} = 0.17\sim 0.27,~
|\phi_1| = 1.2\sim 2.0,~ |\delta\phi|= 0.08\sim 0.10\,,
\end{split}
\end{equation}
for the normal mass ordering, and
\begin{equation}
\begin{split}
&y^0_1 = y^0_2 = 1.2,~y^0_3 = 1.0,~
y_1 = -y_2 = y_3  = y_1^\prime = y_2^\prime = -y_3^\prime = 0.8,\\
&y_1'' = -y_2'' = -y_3'' = 0.84\sim 0.88,\\
&f_\nu^\prime=1.0,~f_\nu=1.05\sim 1.08,~
g_\nu = g_\nu'' =-0.9,~g_\nu^\prime = 1.1 \sim 1.5,\\
&-h_\nu = -h_\nu^\prime = -h_\nu'' = h_\nu''' = h_\nu'''' = 1.3 \sim 1.8,\\
&\frac{s_{1,2}}{\Lambda_F} = 0.18\sim 0.28,~
|\phi_1| = 1.2\sim 2.0,~ |\delta\phi|= 0.08\sim 0.10\,,
\end{split}
\end{equation}
for the inverted mass ordering.
The parameter spaces are required to reproduce the charged lepton mass ratios at the $Z$-boson mass scale \cite{rge}:
\begin{eqnarray}
\frac{m_e}{m_\mu}=4.74\times 10^{-3},~~
\frac{m_\mu}{m_\tau}=5.88\times 10^{-2}\,,
\end{eqnarray}
and to satisfy $1\sigma$ constraints of the oscillation parameters:
\begin{eqnarray}
\frac{\Delta m_{12}^2}{\Delta m_{23}^2}\eqn{=}
\left\{\begin{array}{l}
(2.94\sim 3.35)\times 10^{-2} \hspace{1cm}{\rm for~Normal} \\
(2.97\sim 3.30)\times 10^{-2} \hspace{1cm}{\rm for~Inverted}
\end{array}\right. , \\
\sin^2\theta_{12}^{\rm PMNS}\eqn{=}
\left\{\begin{array}{l}
0.291\sim 0.325 \hspace{1cm}{\rm for~Normal} \\
0.303\sim 0.336 \hspace{1cm}{\rm for~Inverted} 
\end{array}\right. ,\\
\sin^2\theta_{13}^{\rm PMNS}\eqn{=}
\left\{\begin{array}{l}
(2.16\sim 2.66)\times 10^{-2} \hspace{1cm}{\rm for~Normal} \\
(2.23\sim 2.76)\times 10^{-2} \hspace{1cm}{\rm for~Inverted}
\end{array}\right. , \\
\sin^2\theta_{23}^{\rm PMNS}\eqn{=}
\left\{\begin{array}{l}
0.365\sim 0.410 \hspace{1cm}{\rm for~Normal} \\
0.569\sim 0.626 \hspace{1cm}{\rm for~Inverted}
\end{array}\right.\,,
\end{eqnarray}
from Refs. \cite{fogli} (for Normal) and \cite{valle} (for Inverted).
Note that $\theta_{12}^{\rm PMNS}$ can be large owing to the mass degeneracy, but at least $10 \%$ tuning is necessary between $f_\nu$ and $f_\nu^\prime$ in order to reproduce a large $\theta_{23}^{\rm PMNS}$.
Note also that, as discussed below Eq. (\ref{eq:Diag_Ml_R}), the scales of $s_i^{}/\Lambda_F^{}$ are indeed close to that of $\sqrt{m_\mu/m_\tau}$.

Furthermore, in those parameter spaces, the mass of the second generation neutrino and $\langle m_\nu^{} \rangle$ are computed as
\begin{eqnarray}
m_2^\nu = 0.07 \sim 0.08~\ev\,,~~\langle m_\nu^{} \rangle = 0.07 \sim 0.08~\ev\,,
\end{eqnarray}
for both the normal and inverted ordering cases.

\subsection{Scalar potential and spontaneous CP violation}
We discuss the possibility for SCPV in our model.
We expand the singlet scalar fields with its VEVs as
\begin{eqnarray}
S \eqn{=} 
\begin{pmatrix}
S_1\\
S_2
\end{pmatrix}
\rightarrow
\begin{pmatrix}
s_1 e^{i\phi_1} + S_1\\
s_2 e^{i\phi_2} + S_2
\end{pmatrix}\,.
\end{eqnarray}
We write the full scalar potential up to renormalizable level:
\begin{eqnarray}
V \eqn{=} V_{H} + V_S + V_{H S}\,, \\
&&V_{H} =
  \alpha \left| H \right|^2 
+ \beta \left| H \right|^4 \,,\label{eq:Higgs_1}\\
&&V_S = 
  \alpha_S \left| S \right|^2 
+ \alpha_S' {\rm Re} \left[ S^2 \right] 
+ \beta_S^a \left| S \right|^2_{\bf 1} \left| S \right|^2_{\bf 1} 
+ \beta_S^b \left| S \right|^2_{\bf 1'} \left| S \right|^2_{\bf 1'} 
+ \beta_S^c \left| S \right|^2_{\bf 2} \left| S \right|^2_{\bf 2} \notag\\
\eqn{}~~~~~~+ \beta_S' {\rm Re} \left[ S^4 \right] 
+ \gamma_S \left| S \right|^2 {\rm Re} \left[ S^2 \right]\,, \\
&&V_{H S} = 
  \lambda \left| H \right|^2 \left| S \right|^2 
+ \lambda' \left| H \right|^2 {\rm Re} \left[ S^2 \right]\,,
\end{eqnarray}
where the couplings $\beta_S^A \,(A=a,b,c)$ distinguish different combinations of $S_{1,2}$ under the $D_N$ tensor product rules, and all of the couplings are supposed to be real. 
Substituting the VEVs into the potential, we obtain
\begin{eqnarray}
V_{H} \eqn{=} \alpha v^2 + \beta v^4 \,,\\
V_S \eqn{=} 
\alpha_S ( s_1^2 + s_2^2 ) 
+ \alpha_S' ( s_1^2 \cos 2 \phi_1 + s_2^2 \cos 2 \phi_2 ) 
+ \beta_S^a ( s_1^2 + s_2^2 )^2 \notag\\
\eqn{} 
- 4 \beta_S^b s_1^2 s_2^2 \sin^2 ( \phi_1 - \phi_2 ) 
+ \beta_S^c \{ ( s_1^2 - s_2^2 )^2 + 4 s_1^2 s_2^2 \cos^2 ( \phi_1 - \phi_2 ) \} \notag\\
\eqn{}
+ \beta_S' \left[ s_1^4 \cos 4 \phi_1 + s_2^4 \cos 4 \phi_2 + 2 s_1^2 s_2^2 \cos [ 2 ( \phi_1 + \phi_2 ) ] \right] \notag\\
\eqn{}
+ \gamma_S ( s_1^2 + s_2^2 ) ( s_1^2 \cos 2 \phi_1 + s_2^2 \cos 2 \phi_2 ) \,, \\
V_{H S} \eqn{=} 
\lambda v^2 ( s_1^2 + s_2^2 ) 
+ \lambda' v^2 ( s_1^2 \cos 2 \phi_1 + s_2^2 \cos 2 \phi_2 )\,.
\end{eqnarray}
Hereafter, we presume that the singlet scalar fields were completely decoupled from the theory at a high energy scale and investigate only the potential $V_S$.
Then, the minimization conditions are calculated as 
\begin{eqnarray}
\frac{\partial V_S}{\partial \phi_1} \eqn{=} 
  -2s_1^2\left[~\{ \alpha_S' + \gamma_S \left( s_1^2 + s_2^2 \right) \} \sin 2 \phi_1 
+ 2 ( \beta_S^b + \beta_S^c ) s_2^2 \sin \left[ 2 \left( \phi_1 - \phi_2 \right) \right] \right.\notag\\
\eqn{}\left. + 2 \beta_S' \left( s_1^2 \sin 4 \phi_1 + s_2^2 \sin \left[ 2 \left( \phi_1 + \phi_2 \right) \right] \right)~\right]=0\,,\label{cond_al_1}\\
\frac{\partial V_S}{\partial \phi_2} \eqn{=} 
  -2s_2^2\left[~\{ \alpha_S' + \gamma_S \left( s_1^2 + s_2^2 \right) \} \sin 2 \phi_2 
- 2 ( \beta_S^b + \beta_S^c ) s_1^2 \sin \left[ 2 \left( \phi_1 - \phi_2 \right) \right] \right]\notag\\
\eqn{}\left.+ 2 \beta_S' \left( s_2^2 \sin 4 \phi_2 + s_1^2 \sin \left[ 2 \left( \phi_1 + \phi_2 \right) \right] \right)~\right]=0 \,,\label{cond_al_2}\\
\frac{\partial V_S}{\partial s_1} \eqn{=} 
2s_1\left[~\alpha_S + \alpha_S' \cos 2 \phi_1 
+ 2 \beta_S^a ( s_1^2 + s_2^2 ) - 4 \beta_S^b s_2^2 \sin^2 ( \phi_1 - \phi_2) \right.\notag\\
\eqn{} 
+ 2 \beta_S^c \{ s_1^2 - s_2^2 + 2 s_2^2 \cos^2 ( \phi_1 - \phi_2 ) \} 
+ 2 \beta_S' ( s_1^2 \cos 4 \phi_1 + s_2^2 \cos \left[ 2 ( \phi_1 + \phi_2 ) \right] ) \notag\\
\eqn{}\left. 
+ \gamma_S \{s_2^2 \cos 2 \phi_2 + ( 2s_1^2 + s_2^2 ) \cos 2 \phi_1 \}~\right]=0
\,,\label{cond_d1}\\
\frac{\partial V_S}{\partial s_2} \eqn{=} 
2s_2\left[~\alpha_S + \alpha_S' \cos 2 \phi_2 
+ 2 \beta_S^a ( s_1^2 + s_2^2 ) - 4 \beta_S^b s_1^2 \sin^2 ( \phi_1 - \phi_2) \right.\notag\\
\eqn{} 
- 2 \beta_S^c \{ s_1^2 - s_2^2 - 2 s_1^2 \cos^2 ( \phi_1 - \phi_2 ) \} 
+ 2 \beta_S' ( s_2^2 \cos 4 \phi_2 + s_1^2 \cos \left[ 2 ( \phi_1 + \phi_2 ) \right] ) \notag\\
\eqn{} \left.
+ \gamma_S \{ s_1^2 \cos 2 \phi_1 + ( s_1^2 + 2s_2^2 ) \cos 2 \phi_2 \}~\right]=0
\,.\label{cond_d2}
\end{eqnarray}
Let us set $\alpha_S'$, $\beta_S'$ and $\gamma_S$ to zero just for simplicity, then the first two conditions become
\begin{eqnarray}
2(\beta_S^b + \beta_S^c)s_2^2\sin[2(\phi_1 - \phi_2)]=0,~~
2(\beta_S^b + \beta_S^c)s_1^2\sin[2(\phi_1 - \phi_2)]=0\,.
\end{eqnarray}
Suppose $\phi_1 \simeq \phi_2$, which is preferred from a model building point of view, the third and fourth conditions give us
\begin{eqnarray}
s_1^2 + s_2^2 \eqn{=} -\frac{\alpha_S}{2 ( \beta_S^a + \beta_S^c )}\,.
\end{eqnarray}

\section{Neutrinoless double beta decay}
We here discuss testability of the models.
In order to keep generality, we introduce the most-general $Z_2$-breaking terms to Eq. (\ref{eq:Mnd}) by hand and re-parametrize it as
\begin{eqnarray}
\frac{\Lambda_\nu}{v^2} 
(V^\ell_{})^T M^\nu V^\ell_{}
\eqn{=} 
\begin{pmatrix}
f_\nu^{} & 0 &0 \\
0 & f_\nu^{} & \bar{M}^\nu_{23} \\
0 & \bar{M}^\nu_{23} & f_\nu^\prime 
\end{pmatrix}
+
\begin{pmatrix}
\varepsilon_{11}^{} &\varepsilon_{12}^{} &\varepsilon_{13}^{}\\
\varepsilon_{12}^{} & 0 & 0\\
\varepsilon_{13}^{} & 0 & 0 
\end{pmatrix},
\label{eq:pMn}
\end{eqnarray}
where all the parameters are complex, the second term violates the $Z_2$ symmetry, and the terms proportional to $s_i^4/\Lambda_F^4$ are neglected in the first term. 
Here, $\bar{M}^\nu_{23}$ is suppressed with $s_i^2/\Lambda_F^2$, and 
as mentioned just below Eq. (\ref{eq:Diag_Ml_R}) the scales of $s_i^{}/\Lambda_F^{}$ can be estimated from the charged lepton mass ratio as $s_i^{}/\Lambda_F^{} \sim \sqrt{m_\mu^{}/m_\tau^{}}$.
Therefore, one can infer $|\bar{M}^\nu_{23}/f_\nu^\prime| \simeq 0.06$.
The scales of $\varepsilon_{ij}^{}$ are unknown, but they may be at least smaller than $\bar{M}^\nu_{23}$ since $\varepsilon_{ij}^{}$ are responsible for a small $\theta_{13}^{\rm PMNS}$.

We evaluate the effective mass of the neutrinoless double beta decay:
\begin{eqnarray}
\langle m_\nu^{} \rangle =
\left| c_{12}^2c_{13}^2m_1^\nu e^{i\gamma_1^{}} + s_{12}^2c_{13}^2m_2^\nu e^{i\gamma_2^{}} + s_{13}^2e^{-2i\delta}m_3^\nu e^{i\gamma_3^{}} \right|,
\nonumber
\end{eqnarray}
where $s_{ij}^{}$ ($c_{ij}^{}$) is $\sin\theta_{ij}^{\rm PMNS}$ ($\cos\theta_{ij}^{\rm PMNS}$), and $\gamma_i^{}$ denotes the Majorana CP violating phases.
In Eq. (\ref{eq:pMn}), phases of $f_\nu^{}$ and $f_\nu^\prime$ can be absorbed into the right-handed charged leptons as we are considering the diagonal basis of the charged lepton mass matrix.
In view of $f_\nu^{},f_\nu^\prime \gg \bar{M}^\nu_{23},\varepsilon_{ij}^{}$, the eigenvalues should be dominated by $f_\nu^{}$ and $f_\nu^\prime$.
As a result, it is conjectured that the Majorana phases are almost vanishing.
Moreover, as long as we focus on the neutrino mass regions where $m_1^\nu \simeq m_2^\nu$ holds, the third term can be dropped because of $s_{13}^2 \ll 1$.
Given these facts, $\langle m_\nu^{} \rangle$ is approximately given by
\begin{eqnarray}
\langle m_\nu^{} \rangle \simeq m_1^\nu \simeq m_2^\nu.
\end{eqnarray}
\begin{figure}[t]
\begin{center}
\includegraphics*[width=0.7\textwidth]{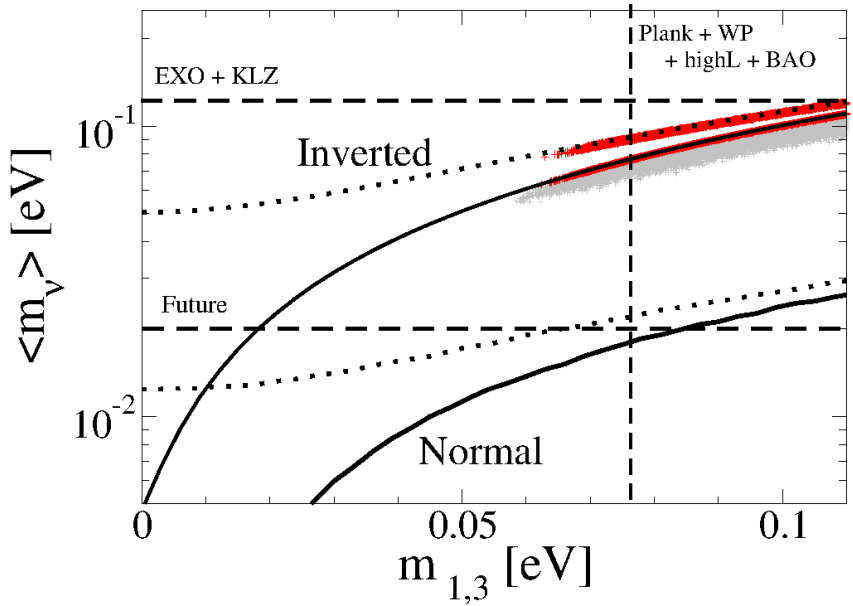}
\end{center}
\vspace{-0.8cm}
\caption{\footnotesize 
The effective mass, $\langle m_\nu^{} \rangle$, of the neutrinoless double beta decay as a function of the lightest neutrino mass, $m_1^\nu$ ($m_3^\nu$) for the normal (inverted) ordering case.
The region surrounded by the solid (dotted) curves corresponds to the normal (inverted) ordering case in the standard $3\nu$ framework, where 3$\sigma$ constraints of the neutrino oscillation parameters from \cite{fogli} are imposed while varying CP phases from $0$ to $2\pi$.
The red regions are favored by the model for $|\bar{M}^\nu_{23}/f_\nu^\prime| > |\varepsilon_{ij}^{}/f_\nu^\prime|$, and the grey region corresponds to the case of $0.5 > |\varepsilon_{ij}^{}/f_\nu^\prime|$ (see text).
The horizontal-dashed lines display the strongest upper bound on $\langle m_\nu^{} \rangle$ from the combined analysis of the EXO \cite{exo} and KamLAND-Zen \cite{kam} experiments, and their expected future bound \cite{neut12}.
The vertical-dashed line represents the $95\%$ C.L. upper bound on the neutrino mass from the Planck data \cite{planck} in combination with a WMAP polarization low-multipole likelihood (WP), the high-resolution CMB data (highL), and constraints from baryon acoustic oscillation (BAO) surveys.
} 
\label{fig:0nbb}
\end{figure}

In Fig. \ref{fig:0nbb}, we numerically calculate $\langle m_\nu^{} \rangle$ while assuming $|\bar{M}^\nu_{23}/f_\nu^\prime| = 0.06\pm 0.04$ and $|\bar{M}^\nu_{23}/f_\nu^\prime| > |\varepsilon_{ij}^{}/f_\nu^\prime|$ in Eq. (\ref{eq:pMn}).
As can be seen, the model favors very narrow regions, and thus it can easily be confirmed or excluded once the lower bounds on $\langle m_\nu^{} \rangle$ and the neutrino mass are available.
Especially, the experimental sensitivity on $\langle m_\nu^{} \rangle$ would reach these regions in the near future.
As we noticed in Sect. III-A, at least $10\%$ tuning is inevitable between $f_\nu^{}$ and $f_\nu^\prime$ in order to reproduce a large $\theta_{23}^{\rm PMNS}$.
Because of this tuning, the model cannot cover the neutrino mass regions in which $m_3^\nu$ is much far from $m_1^\nu \simeq m_2^\nu$.

The above results and conclusions are based on the requirement $f_\nu^{},f_\nu^\prime \gg \bar{M}^\nu_{23},\varepsilon_{ij}^{}$.
For instance, if one relaxes $|\bar{M}^\nu_{23}/f_\nu^\prime| > |\varepsilon_{ij}^{}/f_\nu^\prime|$ into $0.5 > |\varepsilon_{ij}^{}/f_\nu^\prime|$, the favored regions start to broaden as depicted by the grey region in the case of normal mass ordering.
We also note that Fig. \ref{fig:0nbb} is not a prediction since the number of parameters in the mass matrix is enough to reproduce any data.
Our claim is that $\langle m_\nu^{} \rangle$ would be found in such regions soon in the case of $f_\nu^{},f_\nu^\prime \gg \bar{M}^\nu_{23},\varepsilon_{ij}^{}$, which is suggested by the model-building of the partial mass degeneracy.

\section{Summary and Future Works}
Inspired by $\Delta m_{12}^2 \ll \Delta m_{23}^2$, we focus on the neutrino mass regions in which the first two generation neutrinos are quasi degenerate in mass.
In the limit of $\Delta m_{12}^2 = 0$, a Majorana neutrino mass matrix respects a two-dimensional rotation symmetry: the first two generations constitute a doublet representation while the third one acts as a singlet representation.
In the charged fermion sectors, the symmetry results in zero masses for the first two generations, which may be a reasonable first-order approximation to the hierarchical mass spectra of the charged fermions.
Moreover, the small CKM and large PMNS mixings can naturally be understood as a consequence of the Dirac and Majorana natures of fermions, respectively.
We propose a simple model for the lepton sector by means of a $D_N$ flavor symmetry and find that the smallness of the electron mass and $\theta_{13}^{}$ could be explained by approximate conservation of its $Z_2^{}$ subgroup.
We also point out that the model would be tested by the future neutrinoless-double-beta-decay experiments and cosmological observations.
In order for the $Z_2^{}$ symmetry to be slightly broken, we extend the scalar sector so as to acquire complex VEVs.
Consequently, the electron mass and $\theta_{13}^{\rm PMNS}$ turn out to be related with each other via CP violating phases.
The extended model can successfully reproduce the experimental data without imposing unnatural hierarchies among dimensionless couplings.
However, at least $10\%$ tuning between $f_\nu$ and $f_\nu^\prime$ is necessary in order to generate a large atmospheric mixing.

In the present work, we have adopted $D_N^{}$ as our flavor symmetry in order to concentrate of the flavor puzzles of fermions.
It may be challenging to enlarge the symmetry to $O(2)$ while including an associated new gauge boson and gauge anomalies.
The quark sectors should also be included, and we need to check whether the CP phases responsible for the electron mass can simultaneously explain the up- and the down-quark mass with the Dirac phase in the CKM matrix.
Furthermore, we plan to implement the Leptogenesis mechanism 
within a specific neutrino mass generation framework.
These issues will be studied elsewhere.

\section*{ACKNOWLEDGMENTS}
The authors would thank to J. Kubo and J. Heeck for useful discussions and comments.

\appendix
\section{Group theory of $D_N^{}$}
$D_N^{}$ is a group of a discrete rotation, $R$, and parity, $P$, in the two-dimensional plane, and their two-dimensional matrix representations are given by
\begin{eqnarray}
R^q=
\bmx{cc}
\cos\frac{2\pi}{N}q & \sin\frac{2\pi}{N}q \\
-\sin\frac{2\pi}{N}q & \cos\frac{2\pi}{N}q
\emx,~~
P=
\bmx{cc}
1 & 0 \\
0 &-1
\emx,
\end{eqnarray}
where $q$ is integer.
It can be inferred from $R^N=1$ and $P^2=1$ that $D_N^{}$ includes $Z_N$ and $Z_2$ as subgroups and that $q$ acts as a $Z_N$ charge.
There are $2N$ group elements:
\begin{eqnarray*}
{\cal G}=1,~R,~R^2, \cdots, R^{N-1},~P,~PR,~PR^2, \cdots, PR^{N-1},
\end{eqnarray*}
and the number of irreducible representations of $D_N^{}$ is
\begin{eqnarray*}
&&\frac{N-2}{2}~~{\rm doublets~~plus}~~4~~{\rm singlets}~~{\rm for}~~N={\rm even}, \\
&&\frac{N-1}{2}~~{\rm doublets~~plus}~~2~~{\rm singlets}~~{\rm for}~~N={\rm odd}.
\end{eqnarray*}
In what follows, we concentrate on the case of $N={\rm odd}$.
See Ref. \cite{nADFS} for the $N={\rm even}$ case.

The doublets are labeled with the $Z_N^{}$ charge $q$, such as ${\bf 2}_1$, ${\bf 2}_2$, $\cdots$, ${\bf 2}_{(N-1)/2}$ which are transformed by $R$, $R^2$, $\cdots$, $R^{(N-1)/2}$, respectively, under the rotation.
One may think that there exit $N-1$ doublets, but for instance ${\bf 2}_{N-a}$ where $N-a > (N-1)/2$ can be identified with ${\bf 2}_a$, so that there are only $(N-1)/2$ doublets for $N={\rm odd}$.
The two singlets are invariant under the rotation $R$, but one of them changes the sign under the parity $P$; we here define ${\bf 1}\rightarrow {\bf 1}$ and ${\bf 1}^\prime \rightarrow -{\bf 1}^\prime$ under $P$.

The tensor products between the singlets are
\begin{eqnarray}
{\bf 1}^\prime \otimes {\bf 1}^\prime = {\bf 1},~~~~
{\bf 1} \otimes {\bf 1}^\prime = {\bf 1}^\prime ,
\label{eq:cg1}
\end{eqnarray}
and those between the doublets are
\begin{eqnarray}
&&
\begin{array}{ccccccccc}
\bmx{c} x_1 \\ x_2 \emx & \otimes & \bmx{c} y_1 \\ y_2 \emx & = &
(x_1y_1+x_2y_2) & \oplus & (x_1y_2-x_2y_1) & \oplus &
\bmx{c} x_1y_1-x_2y_2 \\ x_1y_2+x_2y_1 \emx \nonumber \\
{\bf 2}_n & \otimes & {\bf 2}_n & = & {\bf 1} & \oplus &
{\bf 1}^\prime & \oplus & {\bf 2}_{2n} 
\end{array},
\\ \nonumber \\
&&
\begin{array}{ccccccc}
\bmx{c} x_1 \\ x_2 \emx & \otimes & \bmx{c} y_1 \\ y_2 \emx & = &
 \bmx{c} x_1y_1+x_2y_2 \\ x_1y_2-x_2y_1 \emx & \oplus &
\bmx{c}  x_1y_1-x_2y_2 \\ x_1y_2+x_2y_1 \emx \label{eq:cg2}\\
{\bf 2}_n & \otimes & {\bf 2}_{m > n} & = & {\bf 2}_{m-n} & \oplus &
 {\bf 2}_{m+n} 
\end{array}
\end{eqnarray}
if $2n \le (N-1)/2$ and $n+m \le (N-1)/2$, while
\begin{eqnarray}
&&
\begin{array}{ccccccccc}
\bmx{c} x_1 \\ x_2 \emx & \otimes & \bmx{c} y_1 \\ y_2 \emx & = &
(x_1y_1+x_2y_2) & \oplus & (x_1y_2-x_2y_1) & \oplus &
\bmx{c} -x_1y_1+x_2y_2 \\ x_1y_2+x_2y_1 \emx \nonumber \\
{\bf 2}_n & \otimes & {\bf 2}_n & = & {\bf 1} & \oplus &
{\bf 1}^\prime & \oplus & {\bf 2}_{N-2n} 
\end{array},
\\ \nonumber \\
&&
\begin{array}{ccccccc}
\bmx{c} x_1 \\ x_2 \emx & \otimes & \bmx{c} y_1 \\ y_2 \emx & = &
 \bmx{c} x_1y_1+x_2y_2 \\ x_1y_2-x_2y_1 \emx & \oplus &
\bmx{c}  -x_1y_1+x_2y_2 \\ x_1y_2+x_2y_1 \emx \label{eq:cg3} \\
{\bf 2}_n & \otimes & {\bf 2}_{m > n} & = & {\bf 2}_{m-n} & \oplus &
 {\bf 2}_{N-(m+n)} 
\end{array}
\end{eqnarray}
if $2n > (N-1)/2$ and $n+m > (N-1)/2$.
Note that Eqs. (\ref{eq:cg1}) and (\ref{eq:cg2}) are the same as those of $O(2)$, whereas Eq. (\ref{eq:cg3}) is used only for $D_N^{}$.

\section{Group theory of $O(2)$}
$O(2)$ is a group of a continuous rotation and parity in the two-dimensional plane:
\begin{eqnarray}
R^q=
\bmx{cc}
\cos\theta q & \sin\theta q \\
-\sin\theta q & \cos\theta q
\emx,~~
P=
\bmx{cc}
1 & 0 \\
0 &-1
\emx,
\end{eqnarray}
where $\theta$ is a continuous parameter, and $q$ behaves like a $U(1)$ charge in this case.
There exist two singlets, ${\bf 1}$ and ${\bf 1}^\prime$ like $D_N^{}$, and an infinite number of doublets. 
The doublets can be labeled with the $U(1)$ charge, such as ${\bf 2}_1$, ${\bf 2}_2, \cdots$.
The tensor products are the same as Eqs. (\ref{eq:cg1}) and (\ref{eq:cg2}) for all doublets.

\end{document}